\documentclass{article}

\usepackage{arxiv}

\usepackage[utf8]{inputenc} 
\usepackage[T1]{fontenc}    
\usepackage{hyperref}       
\usepackage{url}            
\usepackage{booktabs}       
\usepackage{amsfonts}       
\usepackage{amssymb}        
\usepackage{amsmath}        
\usepackage{nicefrac}       
\usepackage{microtype}      
\usepackage{lipsum}		
\usepackage{graphicx}
\usepackage[numbers]{natbib}
\usepackage{doi}
\usepackage{wrapfig}
\usepackage{tikz-cd}
\usepackage{xcolor}

\title{The free Abelian group in R: the {\tt frab} package}

\author{ \href{https://orcid.org/0000-0001-5982-0415}{\includegraphics[width=0.03\textwidth]{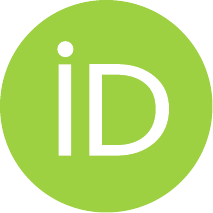}\hspace{1mm}Robin K. S.~Hankin}\thanks{\href{https://academics.aut.ac.nz/robin.hankin}{work};  
\href{https://www.youtube.com/watch?v=JzCX3FqDIOc&list=PL9_n3Tqzq9iWtgD8POJFdnVUCZ_zw6OiB&ab_channel=TrinTragulaGeneralRelativity}{play}} \\
 Auckland University of Technology\\
	\texttt{hankin.robin@gmail.com} \\
}

\hypersetup{
pdftitle={The free Abelian group in R},
pdfsubject={q-bio.NC, q-bio.QM},
pdfauthor={Robin K. S.~Hankin},
pdfkeywords={The free Abelian group, named vectors}
}

\usepackage{Sweave}
\begin{document}
\maketitle

\setlength{\intextsep}{0pt}
\begin{wrapfigure}{r}{0.2\textwidth}
  \begin{center}
\includegraphics[width=1in]{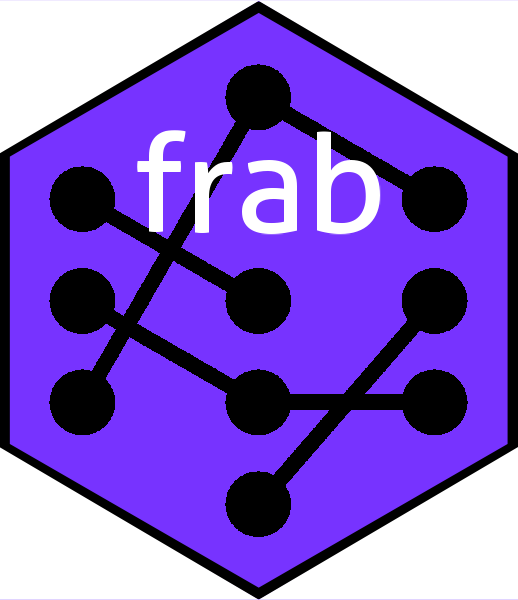}
  \end{center}
\end{wrapfigure}

\begin{abstract}

  In this short article I introduce the {\tt frab} package which
  provides an alternative interpretation of named vectors in the R
  programming language; it is available on CRAN at\\
  \url{https://CRAN.R-project.org/package=frab}.  The underlying
  mathematical object is the free Abelian group.

\end{abstract}

\section{Introduction}

The {\bf Free Abelian Group} is a direct sum of infinite cyclic
groups.  If these cyclic goups are generated by $\left\lbrace
x_i\colon i\in\mathcal{I}\right\rbrace$ for some (finite) index set
$\mathcal{I}$, then the Free Abelian group $F$ will be

$$F=\bigoplus_{i\in\mathcal{I}}\left\langle x_i\right\rangle.$$

From now on we assume that $\left|\mathcal{I}\right| =k < \infty$;
thus the elements of $F$ will be of the form

\begin{equation}\label{formal_form}
g=n_1x_1+n_2x_2+\cdots+n_kx_k
\end{equation}

where $k_i\in\mathbb{Z}$, $1\leqslant i\leqslant k$.  The group
operation (conventionally one uses additive notation) is then defined
by componentwise addition:

$$g=n_1x_1+n_2x_2+\cdots+n_kx_k$$
$$h=r_1x_1+r_2x_2+\cdots+r_kx_k$$

$$h+g=(n_1+r_1)x_1+(n_2+r_2)x_2+\cdots+(n_k+r_k)x_k$$

One can define $F$ formally by starting with a generating set
$X=\left\lbrace x_1,\ldots,x_k\right\rbrace$ of symbols and defining
$F$ as the set of all formal expressions of the form~\ref{formal_form}
under addition as defined above.	 

The Free Abelian group is an interesting and useful mathematical
object.  Here I show how it may be implemented in the R programming
language \citep{rcore2023}.  I also show how a slight natural
generalization (which is convenient in the context of numerical
techniques), may be incorporated.

\section{Package internals}

The package uses the {\tt STL map} class for efficiency.  This class
maps strings (symbols) to doubles; the declaration

\begin{verbatim}
typedef std::map <std::string, double> frab; 
\end{verbatim}

appears in the {\tt src/} package directory.  Such maps are limited
only by memory availability.

\section{The {\tt frab} package in use}

The {\tt frab} package associates a numerical value with each of a set
of arbitrary (character string) symbols.  This is accomplished using
the {\tt STL} {\tt map} class, a container that stores key-value pairs
and allows fast lookup and insertion based on the key.  Here we have
keys as character strings and values are double-precision numbers.

To use the package, it must first be installed and loaded:

\vphantom{f}\\[2\baselineskip]
\begin{minipage}[t]{0.49\textwidth}
\begin{Schunk}
\begin{Sinput}
> install.packages("frab")
\end{Sinput}
\end{Schunk}
\begin{Schunk}
\begin{Sinput}
> library("frab")
\end{Sinput}
\end{Schunk}
\end{minipage}
\hfill
\vrule
\hfill
\begin{minipage}[t]{0.49\textwidth}
\vspace{-\baselineskip}
\color{violet}
\begin{itemize}
\item Function {\tt install.packages()} downloads packages from CRAN
\item Function {\tt library()} loads packages to the current R session
\end{itemize}
\color{black}
\end{minipage}
\vphantom{f}\\[2\baselineskip]

The package uses a single S4 class, {\tt frab}, for which a variety of
methods is defined.  There are several ways to create {\tt frab}
objects, but the most straightforward is to coerce a named vector
using the {\tt frab()} function:

\vphantom{f}\\[2\baselineskip]
\begin{minipage}[t]{0.49\textwidth}
\begin{Schunk}
\begin{Sinput}
> frab(c(z=2,y=7,x=1))
\end{Sinput}
\begin{Soutput}
A frab object with entries
x y z 
1 7 2 
\end{Soutput}
\end{Schunk}
\end{minipage}
\hfill
\vrule
\hfill
\begin{minipage}[t]{0.49\textwidth}
\vspace{-\baselineskip}
\color{violet}
\begin{itemize}
\item Function {\tt frab()} takes a named vector as its single argument
\item It returns an object of class {\tt frab}
\item The elements of the returned {\tt frab} object are reordered; they
appear in an implementation-specific order
\end{itemize}
\color{black}
\end{minipage}
\vphantom{f}\\[2\baselineskip]

Above, see how {\tt frab()} takes a named numeric vector and returns
an object of class {\tt frab}.  It takes the names of its argument,
possibly reordering them, and returns a {\tt frab} object.  Function
{\tt frab()} considers the names of the elements to be the primary
extraction and replacement mechanism.  If the argument has repeated
names, function {\tt frab()} sums them:

\vphantom{f}\\[2\baselineskip]
\begin{minipage}[t]{0.49\textwidth}
\begin{Schunk}
\begin{Sinput}
> frab(c(t=3,q=2,t=4,q=-1,p=6,a=3,t=5))
\end{Sinput}
\begin{Soutput}
A frab object with entries
 a  p  q  t 
 3  6  1 12 
\end{Soutput}
\end{Schunk}
\end{minipage}
\hfill
\vrule
\hfill
\begin{minipage}[t]{0.49\textwidth}
\vspace{-\baselineskip}
\color{violet}
\begin{itemize}
\item Function {\tt frab()} coerces its argument, a named
vector, to an object of class {\tt frab}
\item Element {\tt t}  and {\tt t} are summed, with values $3+5=8$ and 
$3+4+5=12$ respectively
\end{itemize}
\color{black}
\end{minipage}
\vphantom{f}\\[2\baselineskip]

Above we see that the entries for {\tt t} and {\tt q} are summed.
Zero entries are discarded:

\vphantom{f}\\[2\baselineskip]
\begin{minipage}[t]{0.49\textwidth}
\begin{Schunk}
\begin{Sinput}
> frab(c(pear=1,kiwi=0,fig=3,lime=2,fig=-3))
\end{Sinput}
\begin{Soutput}
A frab object with entries
lime pear 
   2    1 
\end{Soutput}
\end{Schunk}
\end{minipage}
\hfill
\vrule
\hfill
\begin{minipage}[t]{0.49\textwidth}
\vspace{-\baselineskip}
\color{violet}
\begin{itemize}
\item Function {\tt frab()} coerces its argument, a named
vector, to an object of class {\tt frab}
\item Element {\tt kiwi} is discarded, having a zero value
\item Element {\tt fig} vanishes, its entries cancelling
\end{itemize}
\color{black}
\end{minipage}
\vphantom{f}\\[2\baselineskip]

Above we see that zero entries are discarded, irrespective of whether
a zero is explicitly given, or repeated values cancel.  However, the
main motivation for using {\tt frab} objects is that they may be
added:

\vphantom{f}\\[2\baselineskip]
\begin{minipage}[t]{0.49\textwidth}
\begin{Schunk}
\begin{Sinput}
> a <- frab(c(x=2,y=1,z=3))
> b <- frab(c(y=3,x=3,u=1))
> a+b
\end{Sinput}
\begin{Soutput}
A frab object with entries
u x y z 
1 5 4 3 
\end{Soutput}
\end{Schunk}
\end{minipage}
\hfill
\vrule
\hfill
\begin{minipage}[t]{0.49\textwidth}
\vspace{-\baselineskip}
\color{violet}
\begin{itemize}
\item Objects {\tt a} and {\tt b} are of class {\tt frab}
\item Their sum is defined in terms of the keys of the summands, not
position
\item Thus, {\tt a+b} has 5 ($=2+3$) for its {\tt x} entry and 4
($=1+3$) for its {\tt y} entry
\item {\tt a+b} has its entries in implementation-specific order, as per
   {\tt disordR} discipline
\item Also, note that {\tt a+b} has length
4, while {\tt a} and {\tt b} have length 3
\end{itemize}
\color{black}
\end{minipage}
\vphantom{f}\\[2\baselineskip]

\section{Extensions of {\tt frab} objects to floating-point values}

The {\tt frab} class is sufficiently flexible to incorporate
floating-point values, although one has to be a little careful with
numerical round-off errors:

\vphantom{f}\\[2\baselineskip]
\begin{minipage}[t]{0.49\textwidth}
\begin{Schunk}
\begin{Sinput}
> x <- frab(c(a=4,u=pi,p=exp(pi)))
> y <- frab(c(p=-exp(pi)/3,u=-pi))
> z <- frab(c(p=-exp(pi)*2/3))
> x+y+z
\end{Sinput}
\begin{Soutput}
A frab object with entries
           a            p 
4.000000e+00 1.776357e-15 
\end{Soutput}
\end{Schunk}
\end{minipage}
\hfill
\vrule
\hfill
\begin{minipage}[t]{0.49\textwidth}
\vspace{-\baselineskip}
\color{violet}
\begin{itemize}
\item Objects {\tt x}, {\tt y} and {\tt z} are of class {\tt frab}
\item Their sum {\tt x+y+z} should have zero entries for {\tt u} and {\tt p}
\item We see the entry for {\tt u} vanishes\ldots
\item \ldots But the entry for {\tt p} is nonzero, being subject to (small)
    numerical roundoff error
\end{itemize}
\color{black}
\end{minipage}
\vphantom{f}\\[2\baselineskip]

\section{The {\tt frab} package and {\tt disordR} discipline}

The {\tt frab} package conforms to disord
discipline~\cite{hankin2022_disordR}.  Here I present some discussion
of the motivation for this design decision.  Consider the following
short R session:

\vphantom{f}\\[2\baselineskip]
\begin{minipage}[t]{0.49\textwidth}
\begin{Schunk}
\begin{Sinput}
(a <- frab(c(x=2,y=1,u=8,z=3,v=5)))
\end{Sinput}
\begin{Soutput}
A frab object with entries
u v x y z 
8 5 2 1 3 
\end{Soutput}
\begin{Sinput}
> a["x"]
\end{Sinput}
\begin{Soutput}
A frab object with entries
x 
2 
\end{Soutput}
\begin{Sinput}
> a[1]
\end{Sinput}
\begin{Soutput}
Error in .local(x, i, j = j, ..., drop):
      not implemented
> 
\end{Soutput}
\end{Schunk}
\end{minipage}
\hfill
\vrule
\hfill
\begin{minipage}[t]{0.49\textwidth}
\vspace{-\baselineskip}
\color{violet}
\begin{itemize}
\item Object {\tt a} is a map from symbols to numeric values
\item The {\tt STL map} class stores value-key pairs in an undefined order
\item Thus, extracting the value for {\tt "x"} is fine,  but because the order is not
defind it makes no sense to extract the ``first" element
\item And attempting to do so results in a disord discipline error
\end{itemize}
\color{black}
\end{minipage}
\vphantom{f}\\[2\baselineskip]

Observe that we cannot dispense with order of the values entirely,
because sometimes I am interested in the vector of keys, or their
values, in isolation.  If we want to work with the names or values of
a {\tt frab} object, then the {\tt disord} print methods are used:

\begin{Schunk}
\begin{Sinput}
> a <- frab(c(x=2,y=1,z=3))
> names(a)
\end{Sinput}
\begin{Soutput}
A disord object with hash 7f95605c74e1dabab41cac277a097c4c67e3306d and elements
[1] "x" "y" "z"
(in some order)
\end{Soutput}
\begin{Sinput}
> values(a)
\end{Sinput}
\begin{Soutput}
A disord object with hash 7f95605c74e1dabab41cac277a097c4c67e3306d and elements
[1] 2 1 3
(in some order)
\end{Soutput}
\end{Schunk}

Above we see that {\tt names(a)} and {\tt values(a)} return {\tt
disord} objects, in this case with the same hash code which indicates
that the objects are consistent with one another in the sense of {\tt
disordR::consitent()}.  These objects may be displayed and
subsequently manipulated, subject to disord discipline:

\vphantom{f}\\[2\baselineskip]
\begin{minipage}[t]{0.49\textwidth}
\begin{Schunk}
\begin{Sinput}
> (a <- frab(c(x=2,y=1,z=3)))
\end{Sinput}
\begin{Soutput}
A frab object with entries
x y z 
2 1 3 
\end{Soutput}
\begin{Sinput}
> names(a) <- toupper(names(a))
> a
\end{Sinput}
\begin{Soutput}
A frab object with entries
X Y Z 
2 1 3 
\end{Soutput}
\end{Schunk}
\end{minipage}
\hfill
\vrule
\hfill
\begin{minipage}[t]{0.49\textwidth}
\vspace{-\baselineskip}
\color{violet}
\begin{itemize}
\item {\tt a} is a {\tt frab} object
\item {\tt names(a)} is a {\tt disord} object as above
\item Replacement methods are defined, in this case {\tt toupper()}
    returns a {\tt disord} object
\item The names of {\tt a} become their uppercase equivalents
\end{itemize}
\color{black}
\end{minipage}
\vphantom{f}\\[2\baselineskip]

Again observe that there is no meaning to the operation ``extract the
first element of {\tt names(a)}", because the elements of {\tt
names(a)}, being a {\tt disord} object, are stored in an
implementation-specific order.  We may manipulate the values of a {\tt
frab} object, if we are careful to be consistent with disord
discipline.  The package includes a number of convenient replacement
idioms:

\vphantom{f}\\[2\baselineskip]
\begin{minipage}[t]{0.49\textwidth}
\begin{Schunk}
\begin{Sinput}
> (a <- frab(c(x=2,y=-1,z=3,p=-4,u=20)))
\end{Sinput}
\begin{Soutput}
A frab object with entries
 p  u  x  y  z 
-4 20  2 -1  3 
\end{Soutput}
\begin{Sinput}
> values(a) <- values(a)^2
> a
\end{Sinput}
\begin{Soutput}
A frab object with entries
  p   u   x   y   z 
 16 400   4   1   9 
\end{Soutput}
\end{Schunk}
\end{minipage}
\hfill
\vrule
\hfill
\begin{minipage}[t]{0.49\textwidth}
\vspace{-\baselineskip}
\color{violet}
\begin{itemize}
\item {\tt a} is a {\tt frab} object
\item We square the {\tt values()} of {\tt a} using the replacement method
\item And object {\tt a} is altered appropriately
\end{itemize}
\color{black}
\end{minipage}
\vphantom{f}\\[2\baselineskip]

Further, we may use the {\tt disindex} class of the {\tt disordR}
package to replace certain values using standard square bracket
replacement idiom:

\vphantom{f}\\[2\baselineskip]
\begin{minipage}[t]{0.49\textwidth}
\begin{Schunk}
\begin{Sinput}
> (a <- frab(c(x=2,y=-1,z=11,p=-4,u=20)))
\end{Sinput}
\begin{Soutput}
A frab object with entries
 p  u  x  y  z 
-4 20  2 -1 11 
\end{Soutput}
\begin{Sinput}
> a[a>10] <- 19
> a
\end{Sinput}
\begin{Soutput}
A frab object with entries
 p  u  x  y  z 
-4 19  2 -1 19 
\end{Soutput}
\end{Schunk}
\end{minipage}
\hfill
\vrule
\hfill
\begin{minipage}[t]{0.49\textwidth}
\vspace{-\baselineskip}
\color{violet}
\begin{itemize}
\item {\tt a} is a {\tt frab} object
\item We set any value exceeding 10 to 19
\item And object {\tt a} is altered appropriately
\end{itemize}
\color{black}
\end{minipage}
\vphantom{f}\\[2\baselineskip]

\appendix
\section{Appendix: Named vectors in R}

A {\em named vector} is a vector with a names attribute; they are a
convenient and useful feature of the R programming language (R Core
Team 2022).  Each element of a named vector is associated with a name
or label.  Objects of the {\tt frab} class bears some resemblance to
named vectors.  However, there are some profound differences:

\subsection{Uniqueness of names}

The names of a named vector are not necessarily unique, unlike those
of a {\tt frab} object.  This has consequences for extraction and
replacement operations.  Consider the following:

\vphantom{f}\\[2\baselineskip]
\begin{minipage}[t]{0.49\textwidth}
\begin{Schunk}
\begin{Sinput}
> (x <- c(a=7,b=4,a=3))
\end{Sinput}
\begin{Soutput}
a b a 
7 4 3 
\end{Soutput}
\begin{Sinput}
> x["a"]
\end{Sinput}
\begin{Soutput}
a 
7 
\end{Soutput}
\end{Schunk}
\end{minipage}
\hfill
\vrule
\hfill
\begin{minipage}[t]{0.49\textwidth}
\vspace{-\baselineskip}
\color{violet}
\begin{itemize}
\item Object {\tt x} is a named vector with three elements.  Both the
first and the third element are named {\tt "a"}
\item This is perfectly OK
\item Extracting element {\tt "a"} returns the {\em first} element
with this name (Technically, it returns a named numeric vector of
length 1)
\end{itemize}
\color{black}
\end{minipage}
\vphantom{f}\\[2\baselineskip]

(we note in passing that double square bracket extraction, as in {\tt
x[["a"]]}, returns the value of the first element with name {\tt a}.

\subsection{Replacement methods for named vectors}

Replacement methods for named vectors is also somewhat problematic:

\vphantom{f}\\[2\baselineskip]
\begin{minipage}[t]{0.49\textwidth}
\begin{Schunk}
\begin{Sinput}
> (x <- c(b=7,a=4,b=3,c=5))
\end{Sinput}
\begin{Soutput}
b a b c 
7 4 3 5 
\end{Soutput}
\begin{Sinput}
> x["a"] <- 100
> x
\end{Sinput}
\begin{Soutput}
  b   a   b   c 
  7 100   3   5 
\end{Soutput}
\end{Schunk}
\end{minipage}
\hfill
\vrule
\hfill
\begin{minipage}[t]{0.49\textwidth}
\vspace{-\baselineskip}
\color{violet}
\begin{itemize}
\item Object {\tt x} is a named vector with four elements.  Both the
first and the third element are named {\tt "a"}
\item Replacing the element {\tt "a"} with 100 behaves as expected:
the element with name {\tt "a"} is returned
\end{itemize}
\color{black}
\end{minipage}
\vphantom{f}\\[2\baselineskip]

This might lead one to believe that replacement of multiple elements
would behave as expected.  But:

\vphantom{f}\\[2\baselineskip]
\begin{minipage}[t]{0.49\textwidth}
\begin{Schunk}
\begin{Sinput}
> x <- c(b=7,a=4,b=3,c=5)
> x["b"] <- 100
> x
\end{Sinput}
\begin{Soutput}
  b   a   b   c 
100   4   3   5 
\end{Soutput}
\end{Schunk}
\end{minipage}
\hfill
\vrule
\hfill
\begin{minipage}[t]{0.49\textwidth}
\vspace{-\baselineskip}
\color{violet}
\begin{itemize}
\item Object {\tt x} is as before
\item Replacing the elements (putatively) indexed with {\tt "b"} [of
which there are two] with {\tt 100} results in only one element
being replaced.
\end{itemize}
\color{black}
\end{minipage}
\vphantom{f}\\[2\baselineskip]

We may also use multiple names for the index in a replacement operation:

\vphantom{f}\\[2\baselineskip]
\begin{minipage}[t]{0.49\textwidth}
\begin{Schunk}
\begin{Sinput}
> x <- c(a=7, b=4, a=3, c=5)
> x[c("a","c")] <- c(100,101)
> x
\end{Sinput}
\begin{Soutput}
  a   b   a   c 
100   4   3 101 
\end{Soutput}
\end{Schunk}
\end{minipage}
\hfill
\vrule
\hfill
\begin{minipage}[t]{0.49\textwidth}
\vspace{-\baselineskip}
\color{violet}
\begin{itemize}
\item Object {\tt x} is as before
\item Replacing the elements (putatively) indexed with {\tt
c("a","c")} with {\tt c(100,101)} replaces the first (but not the
second) of the {\tt "a"} elements, and the {\tt "c"} element, with the
replacement value
\end{itemize}
\color{black}
\end{minipage}
\vphantom{f}\\[2\baselineskip]

\subsection{Addition of named vectors}

Named vectors obey the usual algebraic relations for vectors, although
the details can be unexpected.  Firstly, if nontrivial recycling rules
are applied, the result retains only the names of the longer of the
two addends:

\vphantom{f}\\[2\baselineskip]
\begin{minipage}[t]{0.49\textwidth}
\begin{Schunk}
\begin{Sinput}
> (x <- c(a=7, b=4, a=3, c=5))
\end{Sinput}
\begin{Soutput}
a b a c 
7 4 3 5 
\end{Soutput}
\begin{Sinput}
> x + c(uu=100)
\end{Sinput}
\begin{Soutput}
  a   b   a   c 
107 104 103 105 
\end{Soutput}
\end{Schunk}
\end{minipage}
\hfill
\vrule
\hfill
\begin{minipage}[t]{0.49\textwidth}
\vspace{-\baselineskip}
\color{violet}
\begin{itemize}
\item Object {\tt x} is a standard named vector
\item Adding {\tt c(uu=100)} [a named vector of length 1] to {\tt x} changes the values but not the names of the result
\end{itemize}
\color{black}
\end{minipage}
\vphantom{f}\\[2\baselineskip]

Now with {\tt x} and {\tt y} named vectors of the same length, there
are at least three plausible values that it might give, viz:

\vphantom{f}\\[2\baselineskip]
\begin{minipage}[t]{0.49\textwidth}
\begin{Schunk}
\begin{Sinput}
> x <- c(a=5,b=3,c=4)
> y <- c(b=4,c=2,a=3)
> plausible1 <- c(a=9,b=5,c=7)
> plausible2 <- c(b=9,c=5,a=7)
> plausible1 <- c(a=8,b=7,c=6)
\end{Sinput}
\end{Schunk}
\end{minipage}
\hfill
\vrule
\hfill
\begin{minipage}[t]{0.49\textwidth}
\vspace{-\baselineskip}
\color{violet}
\begin{itemize}
\item Objects {\tt x} and {\tt y} are standard named vectors
\item Three plausible results: {\tt p1 p2 p3}
\item {\tt p1} adds elementwise and assigns names of {\tt x}
\item {\tt p2} adds elementwise and assigns names of {\tt y}
\item {\tt p3} adds namewise
\end{itemize}
\color{black}
\end{minipage}
\vphantom{f}\\[2\baselineskip]

A good case could be made for any of the plausible outcomes above.
However, in standard R idiom, adding two named vectors is equivalent
to stripping the names attribute, performing the addition, then
inserting the names as appropriate.  If the two addends are of equal
length, the names of the first one is given the the result:

\vphantom{f}\\[2\baselineskip]
\begin{minipage}[t]{0.49\textwidth}
\begin{Schunk}
\begin{Sinput}
> x <- c(a=1,b=2,c=3)
> y <- c(c=4,b=1,a=1)
> x+y
\end{Sinput}
\begin{Soutput}
a b c 
5 3 4 
\end{Soutput}
\begin{Sinput}
> y+x
\end{Sinput}
\begin{Soutput}
c b a 
5 3 4 
\end{Soutput}
\end{Schunk}
\end{minipage}
\hfill
\vrule
\hfill
\begin{minipage}[t]{0.49\textwidth}
\vspace{-\baselineskip}
\color{violet}
\begin{itemize}
\item Objects {\tt x} and {\tt y} are  standard named vectors
\item {\tt x+y} and {\tt y+x} have the same values but different names
\item {\tt x+y} inherits the names of {\tt x}
\item {\tt y+x} inherits the names of {\tt y}
\end{itemize}
\color{black}
\end{minipage}
\vphantom{f}\\[2\baselineskip]

If the addends are of incompatible length, a warning is given:

\vphantom{f}\\[2\baselineskip]
\begin{minipage}[t]{0.49\textwidth}
\begin{Schunk}
\begin{Sinput}
> x <- c(a=1,b=2,c=3)
> y <- c(c=4,b=1,a=1,p=4)
> x+y
\end{Sinput}
\begin{Soutput}
c b a p 
5 3 4 5 
Warning message:
In x + y : longer object length is not a
   multiple of shorter object length
> 
\end{Soutput}
\end{Schunk}
\end{minipage}
\hfill
\vrule
\hfill
\begin{minipage}[t]{0.49\textwidth}
\vspace{-\baselineskip}
\color{violet}
\begin{itemize}
\item Objects {\tt x} and {\tt y} are  standard named vectors
\item {\tt x+y} is calculated using standard recycling rules
\item A warning (and optionally an error) is given
\item Names are inherited from the longer of the two addends, here {\tt y}
\end{itemize}
\color{black}
\end{minipage}
\vphantom{f}\\[2\baselineskip]

\section{Tables in R}

Objects of class {\tt table} are created by function {\tt
table::base()}.  Their behaviour is discussed here.  Suppose we have
three tables:

\vphantom{f}\\[2\baselineskip]
\begin{minipage}[t]{0.49\textwidth}
\begin{Schunk}
\begin{Sinput}
> xl <- c("a","a","b","c","d","d","a")
> yl <- c("a","a","b","d","d","d","e")
> zl <- c("a","a","b","d","d","e","f")
> x <- table(xl)
> y <- table(yl)
> z <- table(zl)
> x
\end{Sinput}
\begin{Soutput}
xl
a b c d 
3 1 1 2 
\end{Soutput}
\end{Schunk}
\end{minipage}
\hfill
\vrule
\hfill
\begin{minipage}[t]{0.49\textwidth}
\vspace{-\baselineskip}
\color{violet}
\begin{itemize}
\item Objects {\tt xl yl zl} are character vectors
\item They may be tabulated using {\tt table()}
\item Object {\tt x} is of class {\tt table}
\item Its entries are in alphabetical order
\item The internal structure of {\tt x} is that of an array with named dimensions
\item Objects {\tt y} and {\tt z} are similar
\end{itemize}
\color{black}
\end{minipage}
\vphantom{f}\\[2\baselineskip]

Can we ascribe any meaning to {\tt x+y}?  We attempt standard R semantics:

\vphantom{f}\\[2\baselineskip]
\begin{minipage}[t]{0.49\textwidth}
\begin{Schunk}
\begin{Sinput}
> x
\end{Sinput}
\begin{Soutput}
xl
a b c d 
3 1 1 2 
\end{Soutput}
\begin{Sinput}
> y
\end{Sinput}
\begin{Soutput}
yl
a b d e 
2 1 3 1 
\end{Soutput}
\begin{Sinput}
> x+y
\end{Sinput}
\begin{Soutput}
xl
a b c d 
5 2 4 3 
\end{Soutput}
\end{Schunk}
\end{minipage}
\hfill
\vrule
\hfill
\begin{minipage}[t]{0.49\textwidth}
\vspace{-\baselineskip}
\color{violet}
\begin{itemize}
\item The sum is defined here
\item No error or warning is given
\item The result is clearly incorrect: the entries for {\tt c,d,e}
should be $1+0=1$, $2+3=5$, and $0+1=1$ respectively
\item The result given by R is comparable to the result of adding named vectors
\end{itemize}
\color{black}
\end{minipage}
\vphantom{f}\\[2\baselineskip]

The correct way to add such tables would be by concatenating their
respective data:

\vphantom{f}\\[2\baselineskip]
\begin{minipage}[t]{0.49\textwidth}
\begin{Schunk}
\begin{Sinput}
> x
\end{Sinput}
\begin{Soutput}
xl
a b c d 
3 1 1 2 
\end{Soutput}
\begin{Sinput}
> y
\end{Sinput}
\begin{Soutput}
yl
a b d e 
2 1 3 1 
\end{Soutput}
\begin{Sinput}
> table(c(xl,yl))
\end{Sinput}
\begin{Soutput}
a b c d e 
5 2 1 5 1 
\end{Soutput}
\end{Schunk}
\end{minipage}
\hfill
\vrule
\hfill
\begin{minipage}[t]{0.49\textwidth}
\vspace{-\baselineskip}
\color{violet}
\begin{itemize}
\item Objects {\tt xl,yl} are defined in the previous chunk
\item We may tabulate {\tt c(xl,yl)}
\item The resulting object in essence sums the named entries of the tables {\tt x,y}
\item For example, the entry for {\tt a} is $3+2=5$
\item This is a reasonable interpretation of ``{\tt x+y}"
\item Note that the result is length 5, and that of each table is 4
\end{itemize} 
\color{black}
\end{minipage}
\vphantom{f}\\[2\baselineskip]

If we do not have access to {\tt xl} and {\tt yl} then the only way to
``add" {\tt x} and {\tt y} would be to reconstruct them:

\vphantom{f}\\[2\baselineskip]
\begin{minipage}[t]{0.49\textwidth}
\begin{Schunk}
\begin{Sinput}
> x
\end{Sinput}
\begin{Soutput}
xl
a b c d 
3 1 1 2 
\end{Soutput}
\begin{Sinput}
> y
\end{Sinput}
\begin{Soutput}
yl
a b d e 
2 1 3 1 
\end{Soutput}
\begin{Sinput}
> (xl_rec <- rep(names(x),times=x))
\end{Sinput}
\begin{Soutput}
[1] "a" "a" "a" "b" "c" "d" "d"
\end{Soutput}
\begin{Sinput}
> (yl_rec <- rep(names(y),times=y))
\end{Sinput}
\begin{Soutput}
[1] "a" "a" "b" "d" "d" "d" "e"
\end{Soutput}
\begin{Sinput}
> table(c(xl_rec,yl_rec))
\end{Sinput}
\begin{Soutput}
a b c d e 
5 2 1 5 1 
\end{Soutput}
\end{Schunk}
\end{minipage}
\hfill
\vrule
\hfill
\begin{minipage}[t]{0.49\textwidth}
\vspace{-\baselineskip}
\color{violet}
\begin{itemize}
\item Objects {\tt x,y} are unchanged
\item We may reconstruct {\tt xl} by using {\tt base::rep(...)} with
the {\tt times} argument to form {\tt xl\_rec}
\item Similarly for {\tt yl}
\item And simply tabulate the concatentation {\tt c(xl\_rec,yl\_rec)} to ``add" {\tt x} and {\tt y}
\item The resulting table correctly sums the entries with regard to their labels
\end{itemize} 
\color{black}
\end{minipage}
\vphantom{f}\\[2\baselineskip]

However, this is extremely inefficient, especially if the entries are
large.  And indeed this method will not work for negative or
non-integral entries, although it is sufficiently robust to
accommodate zero entries consistently.

\bibliographystyle{apalike}
\bibliography{frab}

\end{document}